\documentclass[a4paper]{article}
\usepackage{hyperref}
\usepackage{INTERSPEECH2021}
\usepackage{amsmath}
\title{Deep Discriminative Feature Learning for Accent Recognition}
\name{Wei Wang$^1$, Chao Zhang$^2$, Xiaopei Wu$^{3,*}$}
\address{
  $^{1,2,3}$School of Computer Science and Technology, Anhui University, Hefei 230601, China
}
\email{coolephemeroptera@gmail.com, wxp2001@ahu.edu.cn}

\begin{document}

\maketitle
\begin{abstract}
\label{abstract}
  Accent recognition with deep learning framework is a similar work to deep speaker identification, they're both  expected to give the input speech an identifiable representation.
  Compared with the individual-level features learned by speaker identification network, the deep accent recognition work throws a more challenging point that forging group-level accent features for speakers. 
%  Specifically, under the limited training data , the rapid convergence to petty accent classification target in closed-set doesn't mean the good generalized representation be learned, which always lead to overfitting. 
%  Additionally, in most formal social occasions, the standardized pronunciation makes the accent detection harder. 
 In this paper, we borrow and improve the deep speaker identification framework to recognize accents, in detail, we adopt Convolutional Recurrent Neural Network as front-end encoder and integrate local features using Recurrent Neural Network to make an utterance-level accent representation.
 Novelly, to address overfitting, we simply add Connectionist Temporal Classification based speech recognition auxiliary task  during training,  and for ambiguous accent discrimination, we introduce some powerful discriminative loss functions in face recognition works to enhance the discriminative power of accent features. 
 We show that our proposed network with discriminative training method (without data-augment) is significantly ahead of the baseline system on the accent classification track in the Accented English Speech Recognition Challenge 2020, where the loss function \textsl{Circle-Loss} has achieved the best discriminative optimization for accent representation. 
% Moreover, based on the  data-augment methods proposed by the Top-1 team, employing discriminative losses can further improve the recognition accuracy.
%  In feature learning, the state-of-the-art loss function Circle-Loss have achieved the best discriminative optimization for accent representation.

\end{abstract}

\section{Introduction}
\label{introduction}

    Under a particular language, the accent is a learned or behavioral speaking property which can be influenced by social status, education and residence area \cite{RN1}, where the accent we mainly concern is caused by the regional factor. 
    Accent recognition (AR) technologies \cite{biadsy2011automatic,najafian2020automatic,teixeira1996accent,deshpande2005accent,ahmed2019vfnet,winata2020learning}, which can be used to targetedly address accent-related problems or predict a creditable identity for the speaker to make customized service, have attracted extensive attention in recent years. 
    In this paper, we design a deep framework with discriminative feature learning method on the accent classification track of Accented English Speech Recognition Challenge 2020 (AESRC2020) \cite{shi2021accented}, where the English speech accents in data-set derive from 8 countries. Unlike many solutions, which concentrate on  preprocessings (e.g. top-1 scheme \cite{huang2021aispeech} in AESRC2020), on the basis of the Fbank input-feature and the raw data, we focus on the discriminative model optimization to improve accent classification accuracy.
    
    In deep feature learning, AR task is similar to speaker identification (SI) \cite{okabe2018attentive,shon2018frame,xie2019utterance,nagrani2020voxceleb} and language identification (LI) \cite{rangan2020exploiting,safitri2016spoken,madhu2017automatic}, they both want to give the input speech a distinguishable representation. In the modeling method, they could share the same deep paradigm \cite{xie2019utterance,nagrani2020voxceleb}: (i) employing deep neural network (DNN) to extract frame-level feature, (ii) temporal
    integration of frame-level features, and (iii) discriminative feature learning.
    In our work, taking the 2D speech spectrogram as input, we use  Convolutional Recurrent Neural Network (CRNN) \cite{shi2016end} to extract frame-level descriptors, specifically, our CRNN based extractor is composed of ResNet \cite{he2016deep} and Bidirectional GRU \cite{chung2014empirical} (BiGRU)  network, nextly, the calculated local descriptors are integrated into a global utterance-level feature using BiGRU\footnote{Unlike the many-to-many structure of the BiGRU in encoder, here we adopt the many-to-one mode.}.   
    However, to learn the group-level accent feature for many speakers is harder than to learn individual-level speaker feature. Concretely, this difficulty is reflected in:
    \begin{enumerate}
    \item \textsl{Overfitting:} Under the limited training-data, due to the number of accents is far less than the number of speakers and the input speech contains rich and varied signals, fast convergence under this petty classification target is not equivalent to obtain accurate accent representation. Hence, the learned decision path may be inaccurate even though it works perfectly on closed-set.
    \item \textsl{Difficult Accent Detection:} In many formal social situations, speakers tend to adopt standardized pronunciation (e.g. the mandarin in Chinese), this pronunciation will narrow the differences between different accents and make accent recognition harder. In another view, this indistinguishable accent leads to ambiguous embedding representation in deep network.
    \end{enumerate}
    In order to solve the above two problems, we improve the deep paradigm with the following two solutions: (i) For issue 1, we adopt multitask learning (MTL) training method, that is, we introduce Automatic Speech Recognition (ASR) auxiliary task where we simply add Connectionist Temporal Classification (CTC) based ASR decoder onto the front-end encoder during training. (ii) For issue 2, we employ some popular discriminative loss functions \cite{wang2018additive,wang2018cosface,deng2019arcface,sun2020circle} in face recognition (FR) works to strengthen the discriminative power of accent feature. In addition to the state-of-the-art discriminative optimization works in SI \cite{xie2019utterance,nagrani2020voxceleb,cai2018exploring,li2017deep}, we further introduce the  loss: \textsl{Circle-Loss} \cite{sun2020circle}, an improvement for the problem of inflexible optimization and ambiguous convergence status, to achieve better performance in our experiments.

\section{Deep Classification Architecture}

\begin{figure}[ht]
    \caption{The Proposed Deep Accent Recognition Network}
    \label{Fig.1}
    \centering
    \includegraphics[width=\linewidth]{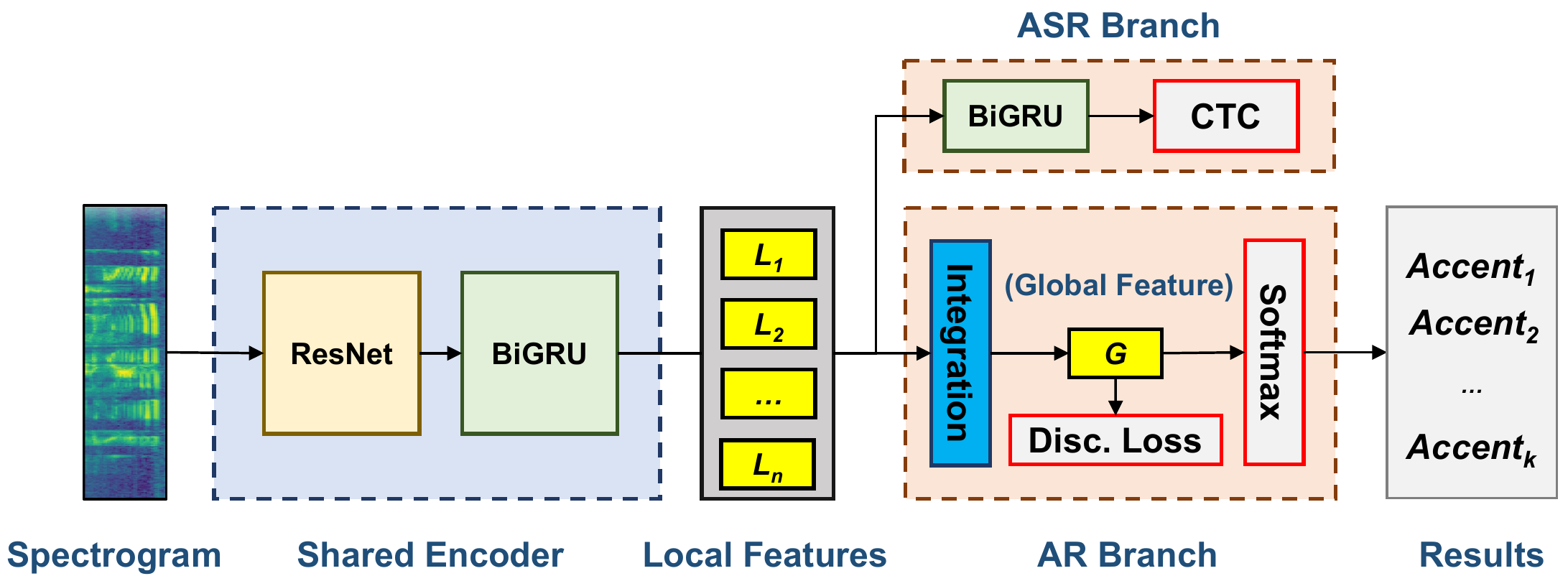}
    \end{figure}

    We borrow and improve the deep SI framework \cite{xie2019utterance,nagrani2020voxceleb} to forge utterance-level representation for accent classification. As \autoref{Fig.1} shows, taking a 2D spectrogram as input, the proposed deep accent recognition network is composed of: 
    (1) a  CRNN based front-end encoder to extract frame-level descriptors, constructed with the ResNet based CNN subpart  and the BiGRU based RNN subpart; 
    (2) a feature integration layer to integrate arbitrary frame-level local features into an  utterance-level global feature vector; 
    (3) a discriminative loss function during training to enhance the discriminative power of global accent feature; 
    (4) a softmax-based classifier, attached to the global feature and give the posterior distribution of accents; 
    The above 2, 3, and 4 components make up the AR branch. To tackle the overfitting issue, we add an ASR branch behind the front-end encoder during training where we simply adopt CTC objective.
    Overall, our model has three outputs (the rectangles with red box in \autoref{Fig.1}) during training,  among which the prediction is given by the  softmax-based classifier. We will give the detailed  modeling explanation in the following subsections. 

\subsection{Front-end Encoder}
\label{front-end-encoder}

    As \autoref{tab1} shows,  after the input spectrogram $\mathcal{X} \in\mathbb{R}^{T \times D \times 1}$ ($T$ denotes the temporal dimension and $D$ denotes the feature dimension ) is feed into the thin 34-layers ResNet (the number of feature maps is cut in half), we obtain the feature maps  $\mathcal{L}_1 \in\mathbb{R}^{(T/2^{5})\times(D/2^{5})\times256}$ (the ResNet has 5 pooling operations overall). 
    Then we could deem tensor $\mathcal{L}_1$ as feature sequence $\mathcal{L}_2 \in\mathbb{R}^{(T \cdot D/2^{10})\times256}$ by combining the time dimension and feature dimension. 
    Next, we can reduce the dimension of descriptors to $\mathcal{H}$ by linear layer and get tensor $\mathcal{L}_3 \in\mathbb{R}^{(T \cdot D/2^{10})\times \mathcal{H}}$. 
    Finally, we adopt BiGRU to further extract sequential feature $\mathcal{L}_4 \in\mathbb{R}^{(T \cdot D/2^{10})\times \mathcal{H}}$.
    In practice, the tensor dimension is rounded up after each pooling, for the pre-pooled dimension $x$, the pooled dimension $P(x)$ is $P(x)=\lceil x/2 \rceil$.
    So, for a  variable-length input speech with $T$ frames and $D$ dimensional acoustic feature,  the amount $C$ of the descriptors we can extract using the proposed encoder is calculated as follow:
    \begin{equation}
    \label{eq1}
        C(T,D)= P(P(P(P(P(T))))) \cdot P(P(P(P(P(D)))))
    \end{equation}
    
        \begin{table}[ht]
        \caption{The construction of CRNN based  font-end encoder }
        \label{tab1}
        \centering
        \resizebox{0.8\linewidth}{12mm}{
        \begin{tabular}{|c|c|c|}
        \hline
        \textbf{Tensor} & \textbf{Layer}           & \textbf{Output Shape}  \\ \hline
        $\mathcal{X}$  & $Spectrogram$                        & $T\times$D$\times1$    \\ \hline
        $\mathcal{L}_1$  & $ResNet$                        & $(T/2^{5})\times(D/2^{5})\times256$    \\ \hline
          $\mathcal{L}_2$  &  $Reshape$                          & $(T \cdot D/2^{10})\times256$   \\ \hline
          $\mathcal{L}_3$ &   $Linear$                              & $(T \cdot D/2^{10})\times \mathcal{H}$                      \\ \hline
          $\mathcal{L}_4$ &   $BiGRU$                           & $(T \cdot D/2^{10})\times \mathcal{H}$                     \\ \hline
        \end{tabular}
        }
        \end{table}

\subsection{Many-to-One Feature Integration}
\label{m2o}

It's a crucial link to merge the variable number of local descriptors caused by the variable-length input into a fixed-length utterance-level embedding vector, that is, we have to transform local descriptors $\mathcal{L}_4$ extracted in \autoref{front-end-encoder} into  global vector $\mathcal{G} \in \mathbb{R}^{\mathcal{H}}$. In this paper, we employ RNN based integration, concretely, we use BiGRU to ingest each descriptor step by step and calculate the last hidden state as an integration result. 

\subsection{CTC-based ASR Objective}
    % As introduced in \ref{introduction}, the rapid convergence under petty classification target does not mean learning a accurate accent representation, which could lead to  dramatical recognition decrease in open-set. 
    % Hence, we will introduce more related clues during training, specially,
    Here, we select CTC-based ASR \cite{watanabe2017hybrid} secondary task to curb overfitting during training. CTC formulation uses L-length letter sequence $O=\{o_l \in \mathcal{U} | l=1,\dots,L\}$ ($\mathcal{U}$ is vocabulary), and framewise letter sequence with 'blank' ($<b>$) symbol: $Z=\{z_t \in \mathcal{U} \cup <b> | t=1,\dots T^{'} \}$. According to  the probabilistic chain rule and conditional independence assumption, for the input $\mathcal{X}$, the CTC objective $p_{ctc}(O|\mathcal{X})$ is factorized as follows:
    \begin{equation}
        \label{eq2}
        p_{ctc}(O|\mathcal{X})=\sum_{Z}\prod_{t}p(z_{t}|z_{t-1},O)p(z_{t}|\mathcal{X})
    \end{equation}
    where $p(z_{t}|z_{t-1},O)$ is transition probability and  $p(z_{t}|\mathcal{X})$ is framewise posterior distribution. In our work, the probability $p(z_{t}|\mathcal{X})$ can be modeled by BiGRU: 
    \begin{gather}
        \label{eq3-4}
        p(z_{t}|\mathcal{X})=Softmax(Lin(h_{t}))\\
        h_{t} = BiGRU_{asr}^{t}(Encoder(\mathcal{X}))
    \end{gather}
    where $Softmax(\cdot)$ is a softmax activation, $Lin(\cdot)$ is a liner layer to convert $\mathcal{H}$ dimensional hidden vector to $(|\mathcal{U}|+1)$ dimensional posterior vector, $Encoder(\cdot)$ is shared front-end extractor and  $BiGRU_{asr}^{t}(\cdot)$ accepts all shared local  descriptors  and outputs hidden vector at $t$ in ASR branch.

\section{Discriminative Feature Learning}
\label{dfl}
    Loss function plays an important role in deep feature learning. 
    Here, we adopt the powerful discriminative losses in FR works to improve the ambiguous accent representation.

\subsection{Softmax Loss}
	\textsl{Softmax} loss, a complex of softmax function and cross entropy loss, is dedicated to making all categories have the largest log likelihood in the probability space.
% 	that is, to ensure that all categories can be classified correctly. 
	Given an utterance-level accent feature $\mathcal{G}_{i}$ with its corresponding label $y_i$, the \textsl{softmax} loss can be formulated as \autoref{eq5}:
	\begin{equation}
	\begin{aligned}
	\mathcal{L} &= \frac{1}{N}\sum_{i}-log\frac{e^{W_{y_i}^T\mathcal{G}_{i}}}{\sum _k e^{W_k^T\mathcal{G}_{i}}}\\
	&= \frac{1}{N}\sum_{i} -log\frac{e^{\Vert W_{y_i} \Vert \Vert  \mathcal{G}_{i} \Vert cos(\theta_{y_i,i})}}{\sum _ke^{\Vert W_{k} \Vert \Vert  \mathcal{G}_{i} \Vert cos(\theta_{k,i})}}
	\end{aligned}
	\label{eq5}
	\end{equation}
	where $N$ denotes the batch size, $W_k$ refers to the $k$-th class weight vector and $\theta _{j,i}$ is the angle between $W_j$ and $\mathcal{G}_i$. But the features trained by this loss is often not good enough for the threshold-based tasks (such as retrieval and verification), that is, the correct classification  does not mean obtaining a metric space with good generalization. 
	
\subsection{CosFace/AM-Softmax}
	In discriminative optimization, we need a reliable metric space, maximizing inter-class variance and minimizing intra-class variance, to improve the generalization in prediction, which is not the target optimized by the \textsl{Softmax} loss function. 
	\cite{wang2018additive} and \cite{wang2018cosface} reformulate the \textsl{Softmax} loss as a cosine loss by L2 normalizing both features and weight vectors to remove radial variations, based on that, an additive cosine margin term is introduced to further maximize the decision margin in the angular space:
	\begin{equation}
	\mathcal{L} = \frac{1}{N}\sum_{i}^{}-log\frac{e^{\gamma \cdot (cos(\theta_{y_i,i})-m)}}{e^{\gamma \cdot (cos(\theta_{y_i,i})-m)}+\sum_{k \neq y_i}e^{cos(\theta_{k,i})}}
	\label{eq6}
	\end{equation}
% 	subject to:
% 	\begin{equation}
% 	W=\frac{W^*}{\left \| W^* \right \|}, \, \mathcal{G}=\frac{\mathcal{G}^*}{\left \| \mathcal{G}^* \right \|}, \, cos(\theta _{j,i}) = W_j^T\mathcal{G}_i
% 	\label{eq7}
% 	\end{equation}
	where 
% 	$W$ and $\mathcal{G}$ are both normalized vector and 
	hyper-parameters $\gamma$ and $m$ are scale factor and margin respectively.
% 	$x_i$ is the $i$-th feature vector corresponding to the ground-truth class of $y_i$, 
% 	the $W_j$ is the weight vector of the  $j$-th class, and $\theta _{j,i}$ is the angle between $W_j$ and $\mathcal{G}_i$, 

\subsection{ArcFace}
	As additive margin as \textsl{Cosface} and \textsl{AM-softmax}, under the normalization on class weights and embedding features, \textsl{ArcFace}\cite{deng2019arcface} moves the margin into the internal of $cos$ operator to optimize the feature space more directly:
	\begin{equation}
	\mathcal{L} = \frac{1}{N}\sum_{i}^{}-log\frac{e^{\gamma \cdot cos(\theta_{y_i,i}+m)}}{e^{\gamma \cdot cos(\theta_{y_i,i}+m)}+\sum_{k \neq y_i}e^{cos(\theta_{k,i})}}
	\label{eq7}
	\end{equation}
	
\subsection{Circle-Loss}
	\cite{sun2020circle} proposes a unified perspective in the deep feature learning including two elemental paradigms: learning with class-level labels  \cite{wang2018additive,wang2018cosface,deng2019arcface} and learning  with pair-wise labels \cite{hoffer2015deep,schroff2015facenet}, which are both aiming  to  maximize  the intra-class similarity $s_p$ and minimize the inter-class similarity $s_n$, concretely, seek to reduce ($s_n-s_p$). 
	Given embedding accent feature $\mathcal{G}$ in the feature space, we assume that there are $K$ intra-class similarity scores and $L$ inter-class similarity scores associated with $\mathcal{G}$. We denote these similarity scores as $\{s_p^i\} (i=1,2,\cdots,K)$ and $\{s_{n}^j\} (j= 1,2,\cdots,L)$, respectively.  A unified loss function can formulated as:
	\begin{equation}
	\begin{aligned}
	\mathcal{L}_{uni} &= log[1+\sum_{i=1}^{K}\sum_{j=1}^{L}e^{\gamma (s_n^j-s_p^i+m)}] \\
	&=log[1+\sum_{j=1}^{L}e^{\gamma (s_n^j+m)}\sum_{i=1}^{K}e^{\gamma (-s_p^i)}]
	\end{aligned}
	\label{eq8}
	\end{equation}
	in which $\gamma$ is a scale factor and $m$ is a margin used for similarity separation. 
	Owing to lacking flexible optimization and ambiguous convergence status in the previous loss functions in reducing ($s_n-s_p$) process, \cite{sun2020circle} proposed \textsl{Circle-Loss} function:
	\begin{equation}
	\begin{aligned}
	\mathcal{L}_{circle} &= log[1+\sum_{j=1}^{L}e^{\gamma \alpha _n^j(s_n^j-\Delta _n)}\sum_{i=1}^{K}e^{-\gamma \alpha _p^i(s_p^i-  \Delta _p)} ] 
	\end{aligned}
	\label{eq9}
	\end{equation}
	where $\Delta _n$ and  $\Delta _p$ are the special margin for $s_n^j$ and $s_p^i$ respectively due to their asymmetric  positions. According to \autoref{eq9}, \autoref{eq10} generalizes $(s_n^j-s_p^i)$ into  $(\alpha _n^j s_n^j - \alpha _p^i s_p^i)$, $\alpha _n^j$ and $\alpha _p^j$ are self-paced weights during  inconsistent gradient descents:
	\begin{equation}
	\left\{\begin{matrix}
	\alpha _p^i = [O_p-s_p^i]+,\\ 
	\alpha _n^j = [s_n^j-O_n]+,
	\end{matrix}\right.
	\label{eq10}
	\end{equation}
	in which $O_n$ and $O_p$ are the optimum for $s_n^j$ and $s_p^j$ respectively, $[\cdot]+$ is the $relu$ activation. Further, \textsl{Circle-Loss} simplifies its hyper-parameters by setting $O_p = 1+m$,  $O_n = -m$, $\Delta _p = 1-m$ and $\Delta _n = m$ with a single margin $m$.
	
\section{Experiment}
\subsection{Dataset}
    We  trained and tested  our  network  on  the AESRC2020 speech data-set \cite{shi2021accented}, an 160 hours open accented English corpus composed of train-set, development-set (dev-set) and test-set for the challenge in accent recognition (track1) and accented speech recognition (track2) with eight national level accents: Chinese (CHN), Indian (IND), Japanese (JPN), Korean (KR), American (US), British (UK), Portuguese (PT), and Russian (RU). Simultaneously, the labels including the accent of the speaker and speech transcript  were offered. 
    The detailed distribution of utterances/speakers (U/S) per national accent was exhibited in \autoref{tab2}. 
    Additionally, the open data-set without accent labels, Librispeech \cite{panayotov2015librispeech}, was allowed to be used in AESRC2020.

	\begin{table*}[ht]
		\centering
		\caption{The detailed distribution of utterances and speakers on the 160 hours data-set of AESRC2020}
		\resizebox{0.8\textwidth}{8mm}{
		\begin{tabular}{ |c|c|c|c|c|c|c|c|c| }
			\hline 
			Data & U/S (CHN) & U/S (IND) & U/S (JAP) & U/S (KR) & U/S (US) & U/S (UK) & U/S (PT) & U/S (RU) \\
			\hline \hline
			train-set & 13583/46 & 13352/38 & 15297/42 & 15807/41 & 16885/64 & 17772/83 & 16343/48 & 14945/37 \\
			\hline 
			dev-set & 1490/4 & 1313/4 & 1488/4 & 1845/4 & 1426/4 & 1751/4 & 1616/4 & 1616/4 \\
			\hline 
			test-set & 1863/- & 1731/- & 1794/- & 1810/- & 2200/- & 1567/- & 1819/- & 1709/- \\
			\hline
		\end{tabular}
		}
		\label{tab2}
	\end{table*}

\subsection{Baseline}
    With Fbank input-feature, the AESRC2020 provided a comparable baseline system realized on EspNet \cite{watanabe2018espnet}, and composed of the transformer \cite{vaswani2017attention} based encoder and Specaugment \cite{park2019specaugment} preprocessing. They integrate temporal descriptors using statistical pooling \cite{shon2018frame}. Additionally, for the overfitting issue, the baseline proposed to initiate the weights of the encoder using a trained ASR encoder-decoder model (ASR-Init.), as a result, the classification accuracy improved significantly.
    
    % We also compared with the preprocessing (T1*) the Top-1 (T1) team focused on  and proposed, where the  preprocessing was composed of Noise Augment \cite{snyder2015musan}, Speed perturb \cite{povey2011kaldi}, Cutting-Splicing \cite{shi2021accented}, Text-to-Speech \cite{tachibana2018efficiently}, Specaugment and ASR-Init.

\subsection{Training}
    We built and trained our model using Keras \cite{gulli2017deep}. In batch training, we extracted 80-dim Fbank acoustic feature and fixed the maxlength of input frames with 1200, according to \autoref{eq1}, we obtained 114 local descriptors finally. 
    Our CTC-based objective adopted BPE \cite{sennrich2015neural} based 1000 subwords and we pre-trained a CTC-based ASR model on AESRC2020 and Librispeech data-set. We restricted the dimension $\mathcal{H}$ of descriptors to 256. We set the scale factor $\gamma$ with 30 for CosFace and ArcFace, and 256 for Circle-Loss. Our MTL loss was defined as: $L = \alpha  L_{ctc} + (1- \alpha) L_{disc.} + \beta  L_{classifier}$, where we assigned $\alpha$ and $\beta$ with 0.4 and 0.01 respectively (We assigned $\beta$ a small value to reduce the impact of softmax classifier on feature space). Additionally, We adopted Adam optimizer \cite{kingma2014adam} and set initial learning rate with 0.01. We executed early-stopping and automatic learning decay trick (the decay factor was 0.3) where the monitor was the accuracy in dev-set. 

\subsection{Results}
    \autoref{tab3} gave the detailed configurations and the accuracy for ours and the baseline. 
    Firstly, without ASR-Init., the baseline systems and our models were both seriously over-fitted, but our accuracy could be improved greatly by adding CTC objective. 
    Next, we compared the scores under the ASR-Init., obviously, the overfitting was improved significantly. Surprisingly, adding CTC objective still possessed a small improvement. In our analysis, the ASR-Init. offered a good starting and MTL maintained the constraint effect throughout the training process. 
    Then, we compared Softmax loss with  discriminative losses (with different margin (m)): CosFace, ArcFace and Circle-Loss. 
    Clearly, the accuracy under discriminative losses was higher than the accuracy under Softmax, and those discriminative results also exceeded the baseline system, among these losses, Circle-Loss with m=0.2 obtained the best result of 68.8\% in the test-set. 
    % Finally, we introduced the preprocessing method proposed by the T1 team, that method won a marvelous result 83.63\% in test-set. 
    % Under Softmax loss, we recur this preprocessing in our work and get 75.5\% accuracy  in test-set, that made a good improvement to our work. Then we replaced Softmax loss by the Circle-Loss with m=0.2 and further improved the accuracy from 75.5\% to 79.3\% in test-set.

	\begin{table*}[ht]
	\centering
	\caption{The results of the baseline, proposed network on the accent classification track of AESRC2020}
	\resizebox{0.8\textwidth}{32mm}{
	\begin{tabular}{ |c|c|c|c|c|c|c|c| }
		\hline
		Exp & PreProc. & Front-end Encoder & ASR Branch & Integration & Disc. Loss & Dev(\%) & Test(\%) \\
		\hline \hline
		\multicolumn{8}{|c|}{\textbf{No ASR-Init.}} \\
		\hline
		Baseline  & Specaugment & Transformer-3L & - & Statistical Pooling  & Softmax & 54.1 & - \\
		Baseline  & Specaugment & Transformer-6L & - & Statistical Pooling & Softmax & 52.2 & - \\
		Baseline  & Specaugment & Transformer-12L & - & Statistical Pooling & Softmax & 47.8 & 33.0 \\
		\hline
		Ours & - &  ResNet-34 + BiGRU & - &  BiGRU & Softmax & 56.2· & 39.0 \\
		Ours & - &  ResNet-34 + BiGRU & CTC & BiGRU & Softmax & 68.5 & 52.7 \\
		\hline \hline
		\multicolumn{8}{|c|}{\textbf{ASR-Init.}} \\
		\hline
		
		Baseline  & Specaugment & Transformer-12L & - & Statistical Pooling & Softmax & 76.1 & 64.9 \\
% 		T1  & - & TDNN & - & - & CosFace & 60.2 & - \\
% 		T1  & - & RES2SETDNN & - & - & CosFace & 57.3 & - \\
		% 			AESRC2020 (top3) & Jsper+Transformer & - & - & - & - & 69.6 \\
		% 			AESRC2020 (top2) & Transformer & - & - & - & - & 72.4 \\
		% 			AESRC2020 (top1) & TDNN & - & - & - & - & 83.6 \\
		\hline
		Ours & - &  ResNet-34 + BiGRU & - & BiGRU & Softmax & 74.8 & 61.4 \\
		Ours & - &  ResNet-34 + BiGRU & CTC & BiGRU & Softmax & 77.3 & 63.9 \\
		Ours & - &  ResNet-34 + BiGRU & CTC &  BiGRU & CosFace(m=0.1) & 79.0 & 65.4 \\
		Ours & - &  ResNet-34 + BiGRU & CTC &  BiGRU & CosFace(m=0.2) & 80.3 & 66.7 \\
		Ours & - &  ResNet-34 + BiGRU & CTC &  BiGRU & CosFace(m=0.3) & 79.3 & 66.6 \\
		Ours & - &  ResNet-34 + BiGRU & CTC &  BiGRU & ArcFace(m=0.1) & 78.6 & 65.4 \\
		Ours & - &  ResNet-34 + BiGRU & CTC &  BiGRU & ArcFace(m=0.2) & 79.4 & 66.7 \\
		Ours & - &  ResNet-34 + BiGRU & CTC &  BiGRU & ArcFace(m=0.3) & 80.2 & 66.5 \\
		Ours & - &  ResNet-34 + BiGRU & CTC &  BiGRU & Circle-Loss(m=0.1) & 81.5 & 68.0 \\
		Ours & - &  ResNet-34 + BiGRU & CTC &  BiGRU & Circle-Loss(m=0.2) & 81.7 & \textbf{68.8} \\
		Ours & - &  ResNet-34 + BiGRU & CTC &  BiGRU & Circle-Loss(m=0.3) & 81.1 & 68.2 \\
% 			\hline
% 			Ours & - & Thin ResNet-34 + BiGRU*1 & CTC &  Avg-Pooling & Circle-Loss(m=0.2) & 81.8 & 67.3\\
% 			Ours & - & Thin ResNet-34 + BiGRU*1 & CTC &  NetVLAD & Circle-Loss(m=0.2) & 81.4 & 67.6\\
% 			Ours & - & Thin ResNet-34 + BiGRU*1 & CTC & GhostVLAD & Circle-Loss(m=0.2) & 81.6 & 67.7\\
		\hline
% 		\textbf{T1}  & \textbf{T1*}. & TDNN \cite{peddinti2015time} & - & unknown & unknown & - & \textbf{83.6} \\
% 		Ours & \textbf{T1*} &  ResNet-34 + BiGRU & CTC &  BiGRU & Softmax & 88.4 & 75.5 \\
% 		Ours & \textbf{T1*} &  ResNet-34 + BiGRU & CTC &  BiGRU & Circle-Loss(m=0.2) & 91.7 & \textbf{79.3} \\

	\end{tabular}
	}
	\label{tab3}
    \end{table*}

    \begin{figure}[htp]
    \centering
    \includegraphics[width=\linewidth]{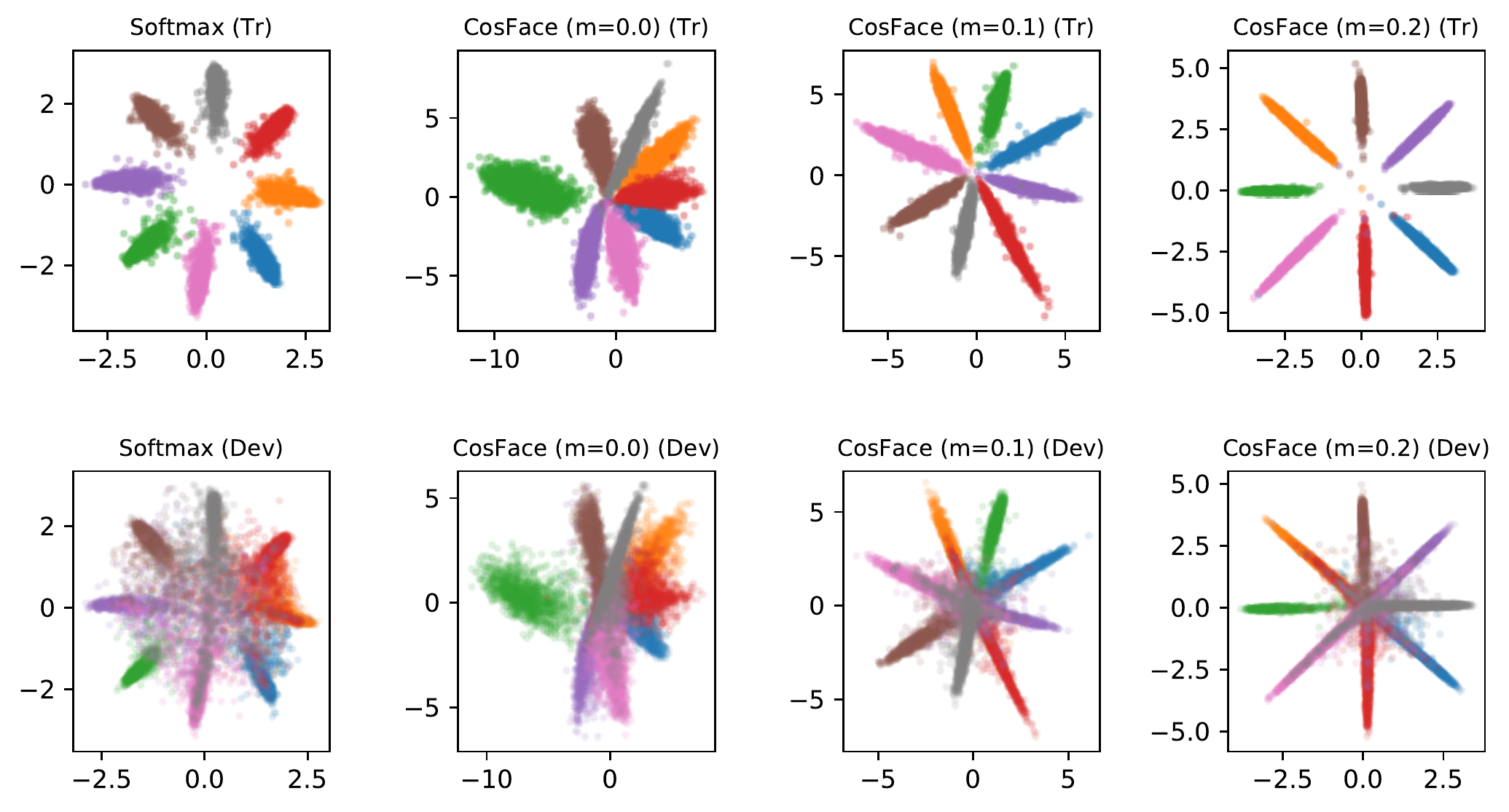}
    \caption{2D embedding accent features trained by Softmax and CosFace (with m= 0.0, 0.1, 0.2) loss functions.}
    \label{Fig.2}
    \end{figure}
    
    \begin{figure}[htp]
    \centering
    \includegraphics[width=\linewidth]{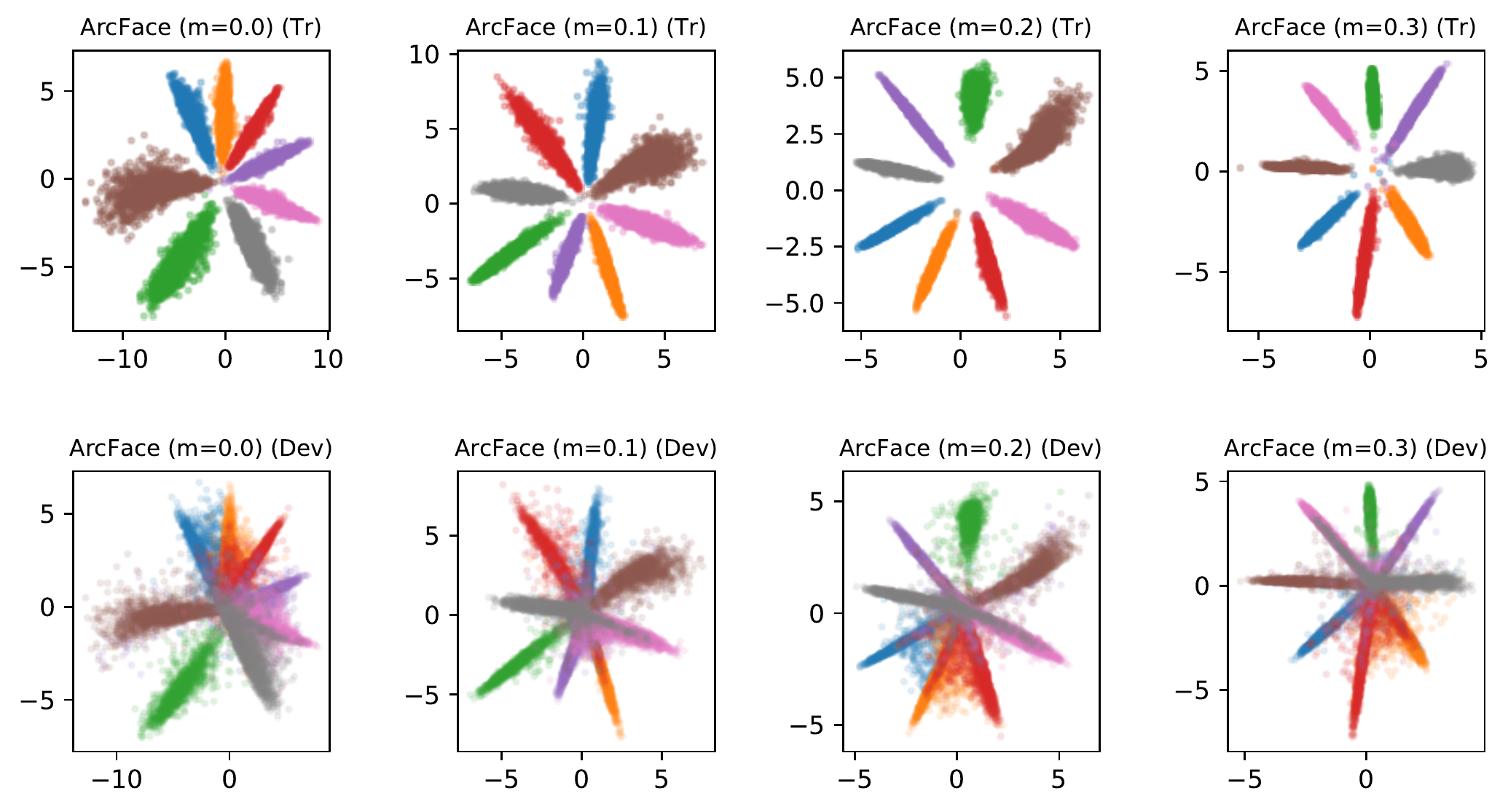}
    \caption{2D embedding accent features trained by ArcFace (with m= 0.0, 0.1, 0.2, 0.3) loss functions.}
    \label{Fig.3}
    \end{figure}
    
    \begin{figure}[htp]
    \centering
    \includegraphics[width=\linewidth]{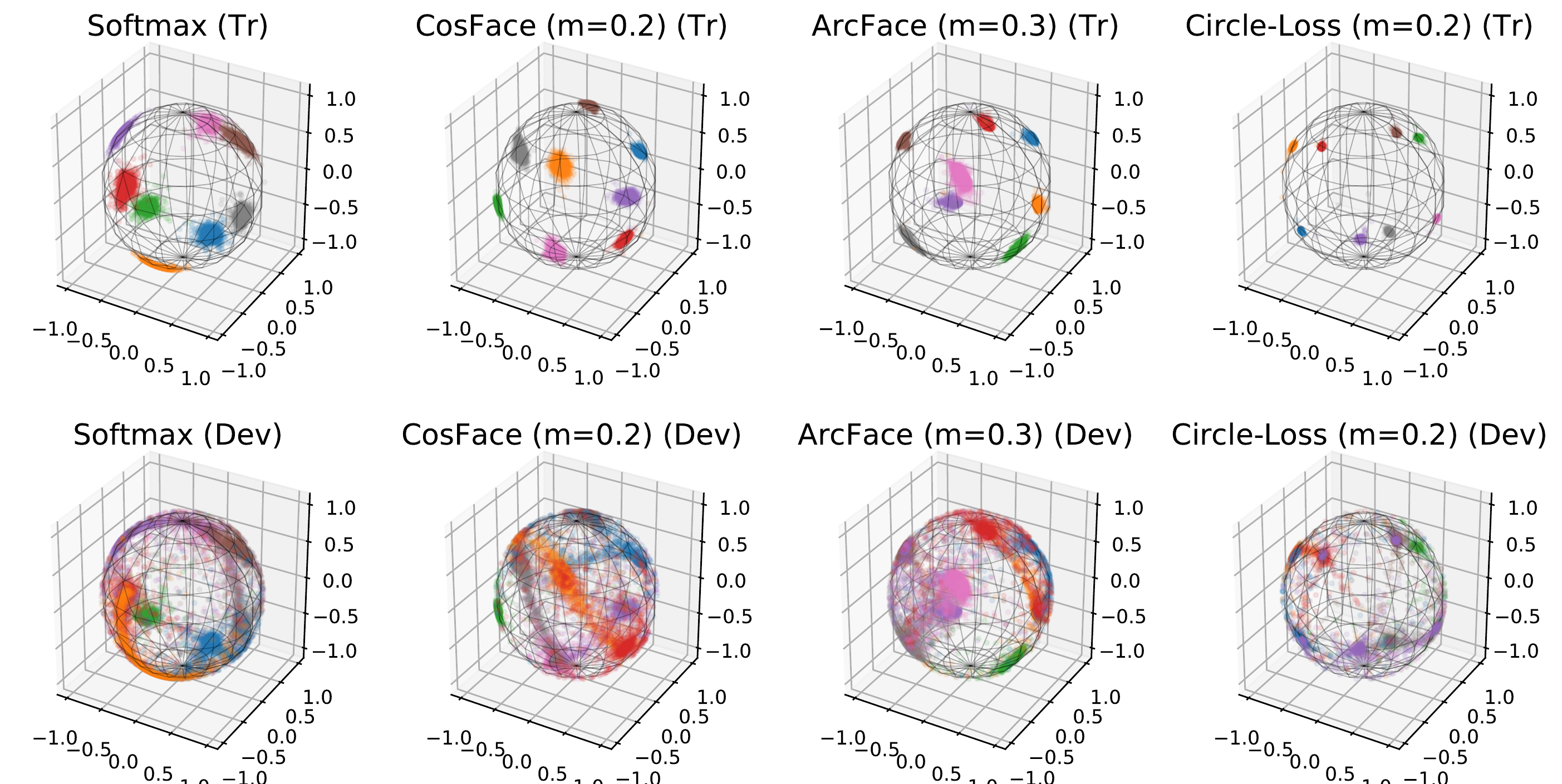}
    \caption{3D embedding unit accent features trained by Softmax, CosFace (with m=0.2), ArcFace (with m=0.3) and Circle-Loss (with m=0.2) loss functions.}
    \label{Fig.4}
    \end{figure}

    In order to better explain the discriminative learning effect, we provided a visual interpretation in low-dimensional feature space, that was, we compacted the embedding accent feature $\mathcal{G}$ into 2D/3D bottleneck feature.  
    \autoref{Fig.2} and \autoref{Fig.3} showed the 2D accent feature distributions of Softmax, CosFace and ArcFace, different colors indicated different regional accents and the first row recorded the train-set and the second row recorded the dev-set, we could observe that Softmax based feature learning led to many ambiguous representations in dev-set, on the contrary, CosFace and ArcFace tightened each category features into a compact space, with the increasing of cosine/angular margin, the compact effect became more significant and we could see a more distinguishable feature distribution in dev-set.
    \autoref{Fig.4} showed the 3D normalized accent feature distributions of Softmax, CosFace, ArcFace and Circle-Loss. Among those losses, Circle-loss made features more compact due to its improved method for flexible optimization and definite convergence status.

    % \autoref{Fig.2} gave the distributions of 2D embedding accent vectors using the Softmax loss and the CosFace loss with m=0.0,0.1,0.2, where the first row recorded the 2D accent representations of train-set and the second row recorded the 2D accent representations of dev-set. Softmax based feature learning did not well in giving the sample in the wild a discriminative identity, which caused many ambiguous representations in dev-set. On the contrary, CosFace loss tightened the features of each category into a compact space, with the increasing of cosine margin, the compact effect became more significant and we could see a more distinguishable feature distribution in dev-set, this phenomenon was also showed in \autoref{Fig.3} (ArcFace with m=0.0,0.1,0.2,0.3). 
    
    % \begin{figure}[t]
    % \centering
    % \includegraphics[width=\linewidth]{LaTeX/transfer.pdf}
    % \caption{Schematic diagram of speech production.}
    % \label{fig:speech_production}
    % \end{figure}

\section{Conclusions}
    In this paper, we borrow the deep paradigm in speaker identification works to propose a deep accent recognition network. Novelly, we employ CTC-based ASR auxiliary task to solve overfitting problem and follow the development of discriminative feature learning in face recognition area to solve ambiguous accent representation. In the AESRC2020 accent recognition track, our proposed models with discriminative optimization both have achieved better scores than the baseline system, and Circle-Loss performs best in the feature learning. 
    Convincingly, some  preprocessings \cite{huang2021aispeech} achieve miraculous scores in AESRC2020, we hope to research with that and explore better scheme  for accent recognition in future work.

\bibliographystyle{IEEEtran}
\bibliography{main}
\end{document}